\documentclass[traditabstract]{aa}
\bibpunct{(}{)}{;}{a}{}{,} 
\usepackage{graphicx}
\usepackage{mathrsfs}
\usepackage[varg]{txfonts}
\usepackage[T1]{fontenc}
\usepackage{color}
\usepackage{stfloats}

\begin{document}

\title{Evidence of a diffuse, extended continuum source in quasars from the relative sizes of the broad line region and the UV-optical continuum source measured with microlensing}
\author{Damien Hutsemékers\inst{1,}\thanks{Research Director F.R.S.-FNRS}
   \and Dominique Sluse\inst{1}
       }
\institute{
    Institut d'Astrophysique et de G\'eophysique,
    Universit\'e de Li\`ege, All\'ee du 6 Ao\^ut 19c, B5c,
    4000 Li\`ege, Belgium
    }
%
%
\titlerunning{Diffuse, extended continuum source in quasars from microlensing} 
\authorrunning{D. Hutsem\'ekers et al.}
\abstract{Microlensing by stars in the lens galaxy of a gravitationally lensed quasar is a phenomenon that can selectively magnify quasar subregions, producing observable changes in the continuum brightness or distortions in the emission line profiles. Hence, microlensing allows us to probe the inner quasar regions. In this paper, we report measurements of the ratio of the broad emission line region (BLR) radius to the continuum source radius in eight lensed quasars, for the \ion{C}{iv}, \ion{Mg}{ii}, and H$\alpha$ emission lines and their respective underlying continua at $\lambda\lambda$ 1550~\AA, 2800~\AA, and 6563~\AA. The microlensing-induced line profile distortions and continuum magnifications were observed in the same single-epoch datasets, and simultaneously compared with microlensing simulations. We found that, on average, the inner radius of the BLR starts at the end of the UV-optical continuum source, independently of the line ionization and the wavelength of the continuum.  The half-light radius of the BLR is, on average, a factor of six larger than the half-light radius of the continuum source, independently of the quasar's bolometric luminosity. We also found a correlation between the BLR radius and the continuum source radius, supporting the idea that the dominant contribution to the UV-optical continuum may come from the BLR itself. Our results independently confirm the results of reverberation mapping studies, and extend them to higher-redshift, higher-luminosity quasars. 
}
\keywords{Gravitational lensing -- Quasars: general -- Quasars:
emission lines}
\maketitle
%
%
%

\section{Introduction}
\label{sec:intro}

The rest-frame UV-optical spectrum of active galactic nuclei (AGNs) and their luminous counterparts, quasars\footnote{In the following, we use the terms AGN and quasar interchangeably.}, is characterized by the presence of a strong, blue continuum, the so-called big blue bump, together with broad emission lines (BELs) of low and high ionization. While it is well established that the AGN phenomenon is powered by the accretion of matter onto a supermassive black hole, the origin of the UV-optical continuum and the properties of the broad emission line region (BLR) are still unclear. Early models assumed that the UV-optical continuum originates in a standard Shakura-Sunyaev thin accretion disk, at a typical distance of one light-day from the supermassive black hole, and that it irradiates a roughly spherically symmetric BLR, two to three orders of magnitude larger based on simple photoionization arguments \citep[e.g.,][]{1995Urry,1997Peterson,2013Netzer}.

By measuring the time delay between the brightness variations in the continuum source and the distant BLR, reverberation mapping (RM) has provided the first reliable estimates of the BLR radius. Based on a statistically significant sample of 19 AGNs, \citet{1999Wandel} reported BLR radii in the range of 1$-$100 light-days, which was significantly smaller than what was expected from early models. Since then, these results have been confirmed for hundreds of quasars \citep[e.g.,][for a review]{2021Cackett}. On the other hand, using microlensing from stars in intervening galaxies as a gravitational telescope, the size of the UV-optical continuum source was found to be a factor of 3$-$4 larger than what was expected from the standard accretion disk model \citep{2010Morgan,2011Blackburne}. This discrepancy was confirmed by intensive wavelength-dependent RM studies of the continuum from the far-UV to the near-IR \citep[e.g.,][]{2016Fausnaugh,2018Cackett,2020Cackett}, making the sizes of the BLR and the continuum source much less different than previously thought. In particular, some of the most comprehensive monitoring campaigns to date have shown that the optical ($\lambda > $ 5000 \AA, rest-frame) continuum source is comparable in size to the inner, highly ionized BLR \citep{2016Fausnaugh,2021Kara}. Although limited to a few local AGNs, these results have profound implications for the interplay between the accretion disk, the UV-optical continuum-emitting source, and the BLR, but they also complicate the interpretation of the RM time lags, since the continuum can no longer be considered as emitted by a point-like source \citep{2017Pei}.

Microlensing by stars in the lens galaxy of a gravitationally lensed quasar is a phenomenon that allows us to probe the properties of the inner regions of the quasar in a different way. Microlensing can selectively magnify quasar subregions, mostly depending on their size, producing changes in the continuum brightness or deformations of the emission line profiles in some images of the lensed quasar. By comparing microlensing-induced variations with simulations, the size of the continuum source and the BLR can be derived \citep[e.g.,][for a recent review]{2024Vernardos}. Compared to RM, which requires intensive monitoring campaigns, microlensing has the advantage that single epoch spectra of the different images of a gravitationally lensed system may be sufficient to derive size estimates. On the other hand, only lensed quasars can be used and a clear microlensing effect must be observed in one of the images.

Recently, we used the microlensing-induced BEL profile distortions observed in single-epoch spectra of eight lensed quasars to constrain the geometry, kinematics, and size of the \ion{C}{iv}, \ion{Mg}{ii}, or H$\alpha$ BLRs \citep{2024Hutsemekers,2024bHutsemekers}. The method not only compares the observed line profile distortions with simulations, but also includes the magnification of the underlying continuum measured on the same spectra, so that the relative sizes of the BLR and the continuum source can be estimated simultaneously. In this paper, we report the measurement of the ratio of the BLR radius to the continuum source radius, for the eight lensed quasars, noting that these quasars have higher redshifts and luminosities than the AGNs targeted by the intensive RM campaigns. In Sect.~\ref{sec:method}, we briefly summarize the method. The results are presented in Sect.~\ref{sec:ratios}, interpreted in Sect.~\ref{sec:discussion}, and summarized in the last section.

\section{Microlensing analysis and simulations}
\label{sec:method}

The analysis of the observed microlensing signals and the comparison with simulations to derive the properties of the BLR are described in \cite{2023Hutsemekers,2024Hutsemekers,2024bHutsemekers}. The models of the BLR and the continuum source are detailed in \cite{2017Braibant}. The probabilistic analysis is developed in \cite{2019Hutsemekers}. Here, we briefly summarize the method. 

We considered eight lensed quasars (Table~\ref{tab:targets}) in which significant line profile deformations due to microlensing were observed in either the \ion{C}{iv} $\lambda$1550, \ion{Mg}{ii} $\lambda$2800, or H$\alpha$ BEL. For each object and line, the distorted line profile observed in the microlensed image was compared to the line profile simultaneously observed in a non-microlensed image to extract, on a velocity scale, the microlensing magnification profile, $\mu(v)$, binned into 20 spectral elements.  At the same time, we evaluated the magnification of the continuum underlying the emission line, $\mu^{cont}$. This constitutes the set of 21 observables to be compared with the simulations.

To simulate the effect of microlensing, we built a model that contains a continuum source and a BLR. The continuum source is modeled by a disk of constant surface brightness of radius $r_{\text s}$. For the BLR models, we considered a rotating Keplerian disk, a biconical, radially accelerated polar wind, and a radially accelerated equatorial wind. We assumed that the emissivity can be expressed as $\varepsilon = \varepsilon_0 \, (r_{\text{in}}/r)^q$, where $r_{\text{in}}$ is the inner radius of the BLR model and $q$ = 3 or 1.5, so that it either decreases sharply with increasing radius, or does so more slowly \citep[see][for details]{2017Braibant}. Using a radiative transfer code, we generated 20 monochromatic (isovelocity) BLR images corresponding to 20 spectral bins in the line profile.

We then computed a magnification map specific to the microlensed image in each lensed system. Distorted line profiles were obtained by convolving, for a given value of $r_{\text{in}}$, the magnification map with the monochromatic images of the BLR. The continuum emitting region was similarly convolved for a given value of $r_{\text s}$.  Simulated $\mu(v)$ and $\mu^{cont}$ were then obtained for each position of the model on the magnification maps. We emphasize that the simulated $\mu(v)$ and $\mu^{cont}$ were not computed independently; since the continuum source and the BLR were attached in the models, only $\mu(v)$ and $\mu^{cont}$ obtained from the same positions on the magnification maps were considered. This process was repeated for a series of values of the radii $r_{\text{in}}$ and $r_{\text s}$ (typically 15 values each\footnote{In some of our papers, the first ones, we only considered models for which $r_{\text{in}} \geq r_{\text s}$ to save computing time. This condition was found to have little effect on the determination of the BLR properties. However, it is clear that it had to be relaxed for the present study.}, approximately 225 combinations), four inclinations of the whole system with respect to the line of sight, two values of $q$, and seven orientations of the magnification map. Finally, the likelihood that the simulations reproduce the 21 observables was computed for each set of parameters.

\section{Ratios of the BLR to continuum source radii}
\label{sec:ratios}

\begin{figure*}[t]
\centering
\resizebox{0.9\hsize}{!}{\includegraphics*{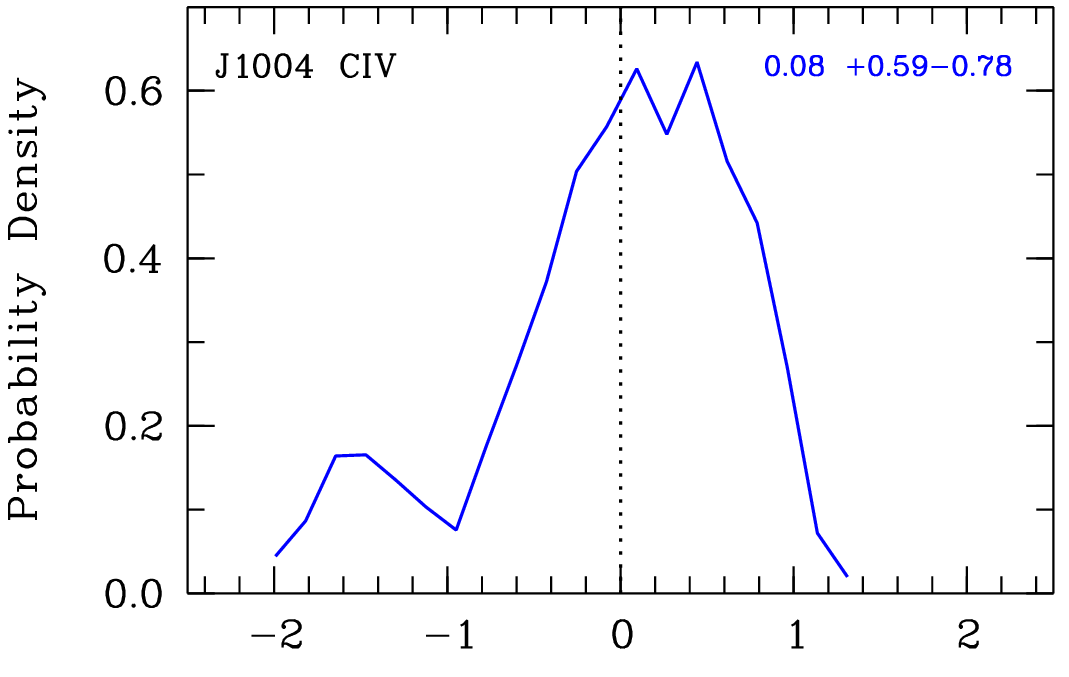}\includegraphics*{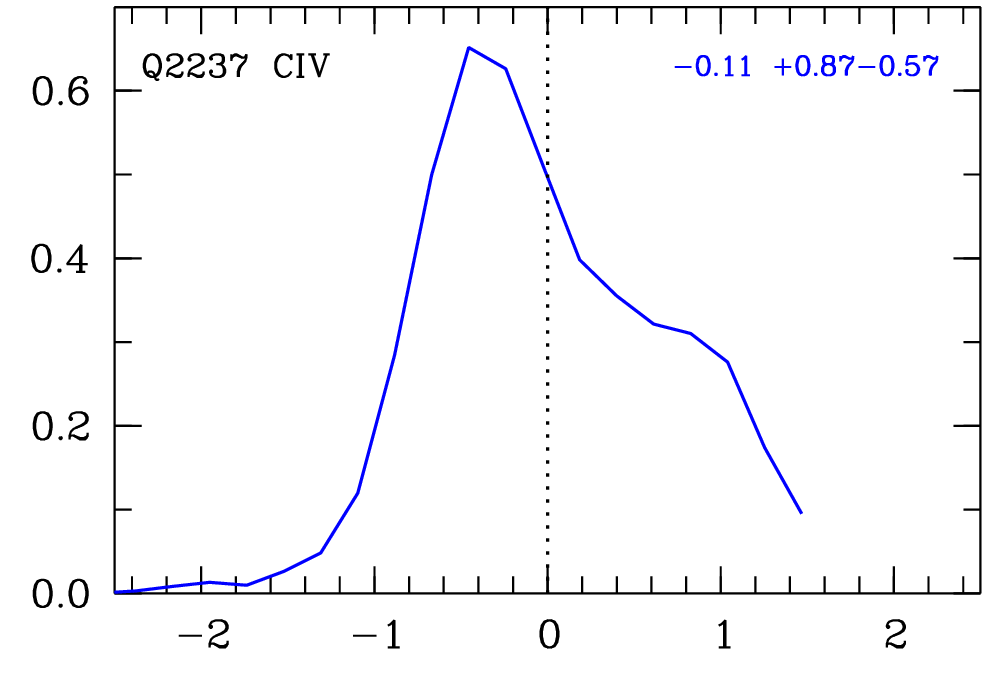}\includegraphics*{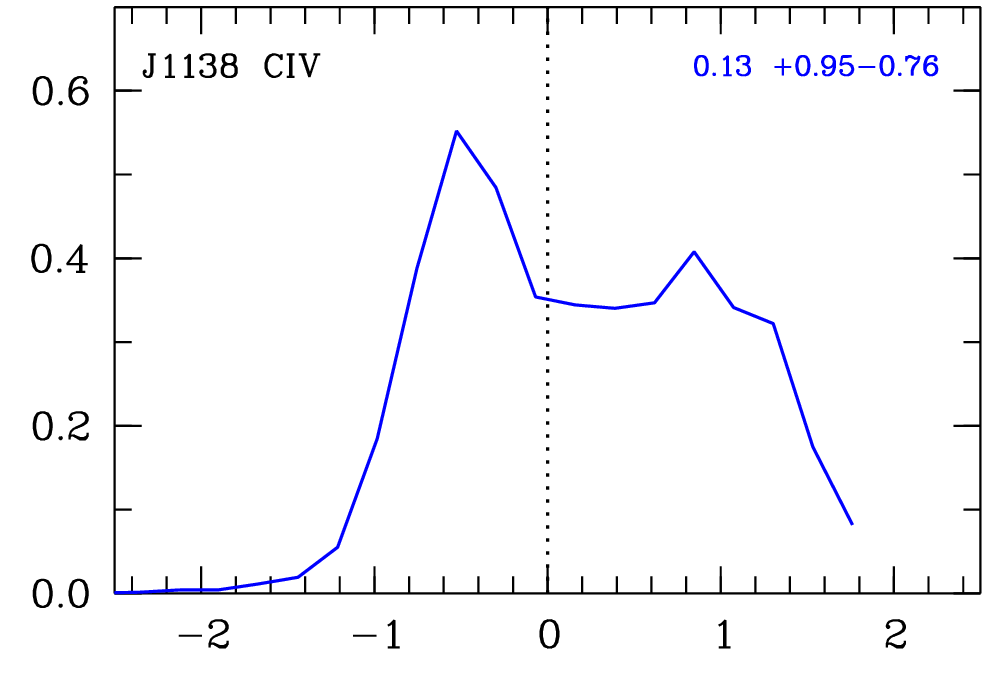}}\\
\resizebox{0.9\hsize}{!}{\includegraphics*{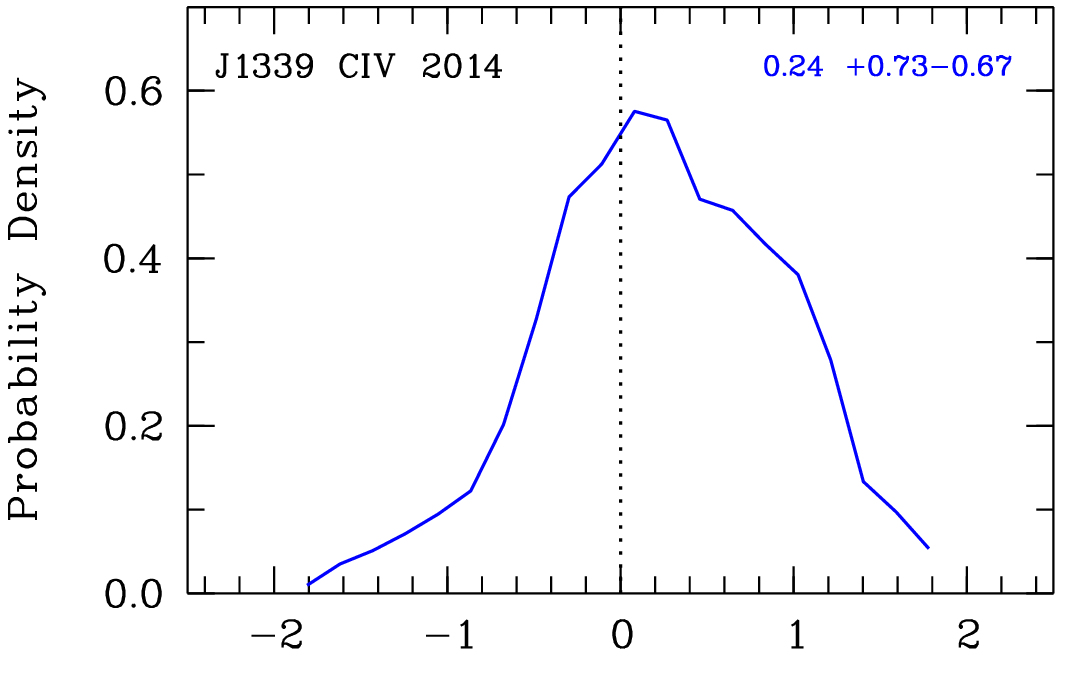}\includegraphics*{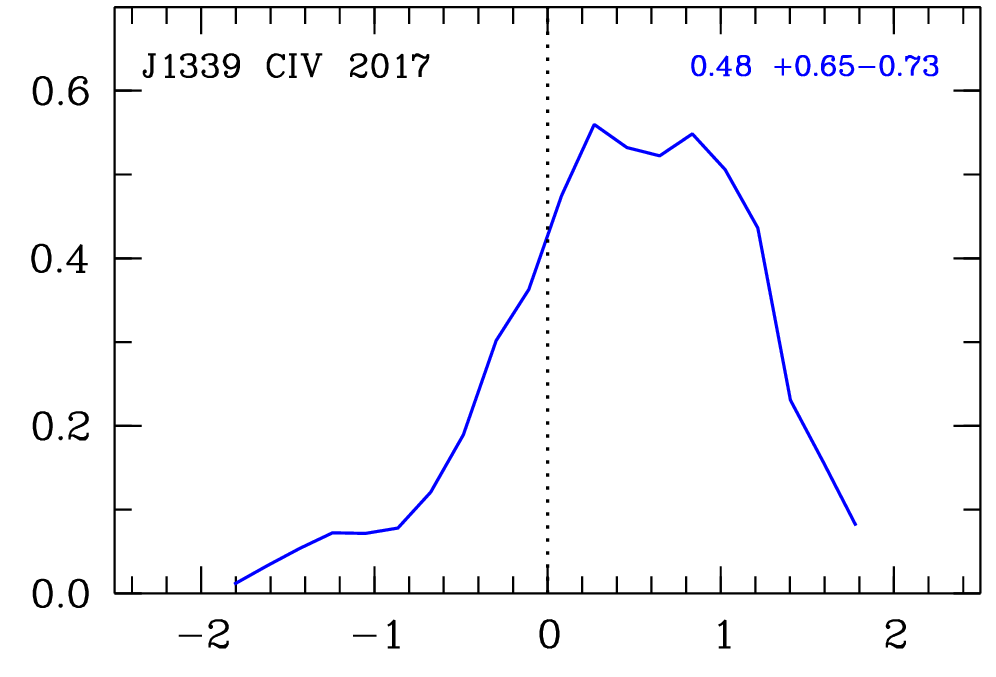}\includegraphics*{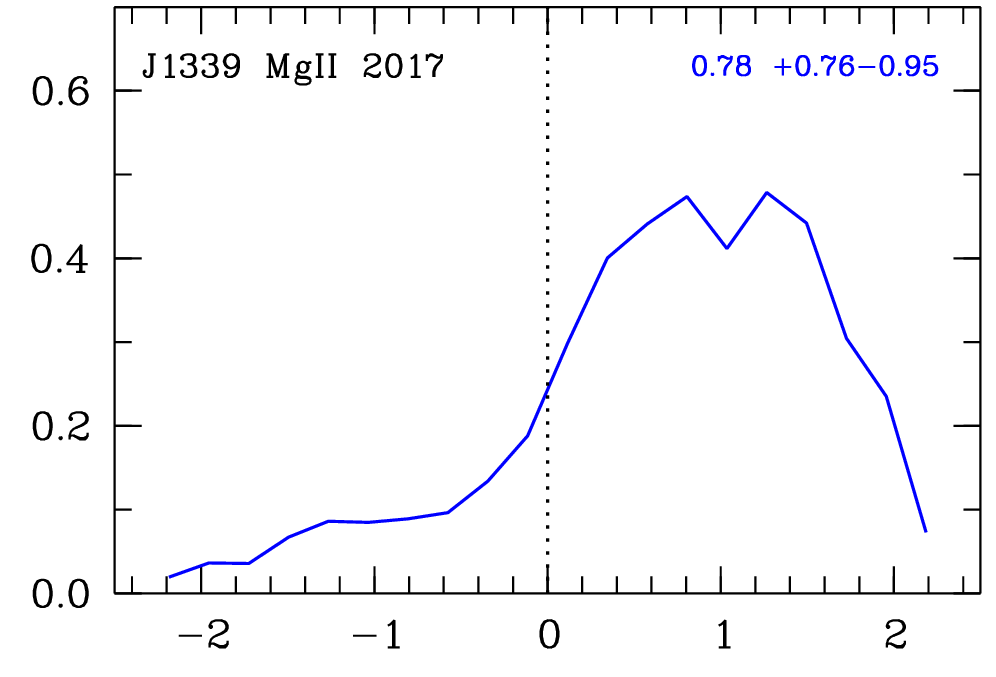}}\\
\resizebox{0.9\hsize}{!}{\includegraphics*{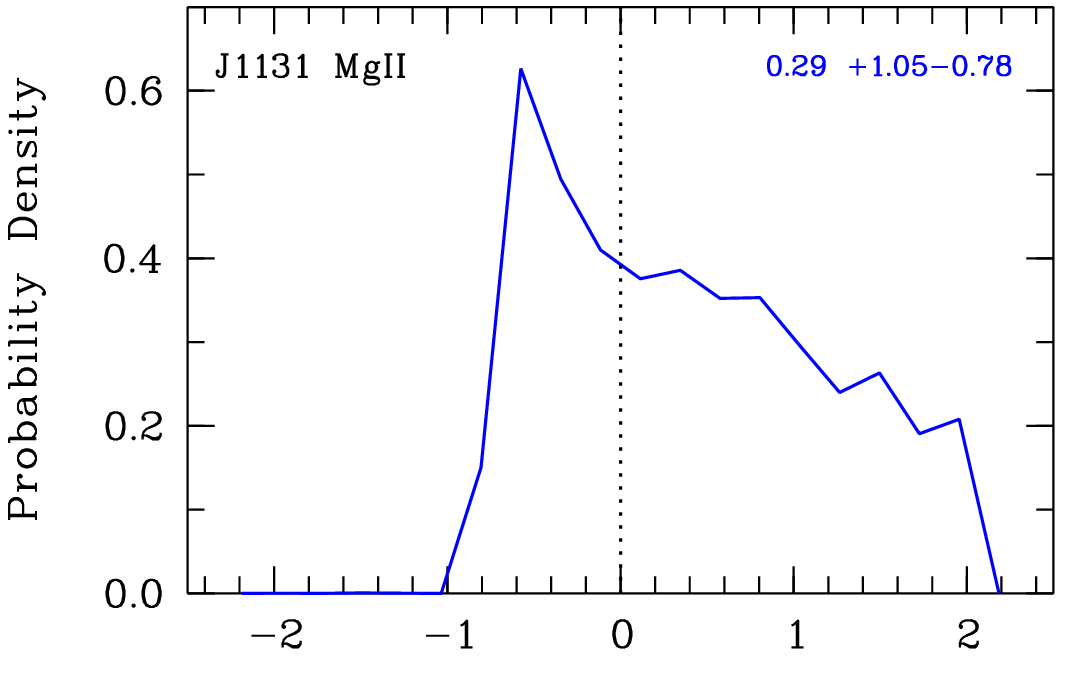}\includegraphics*{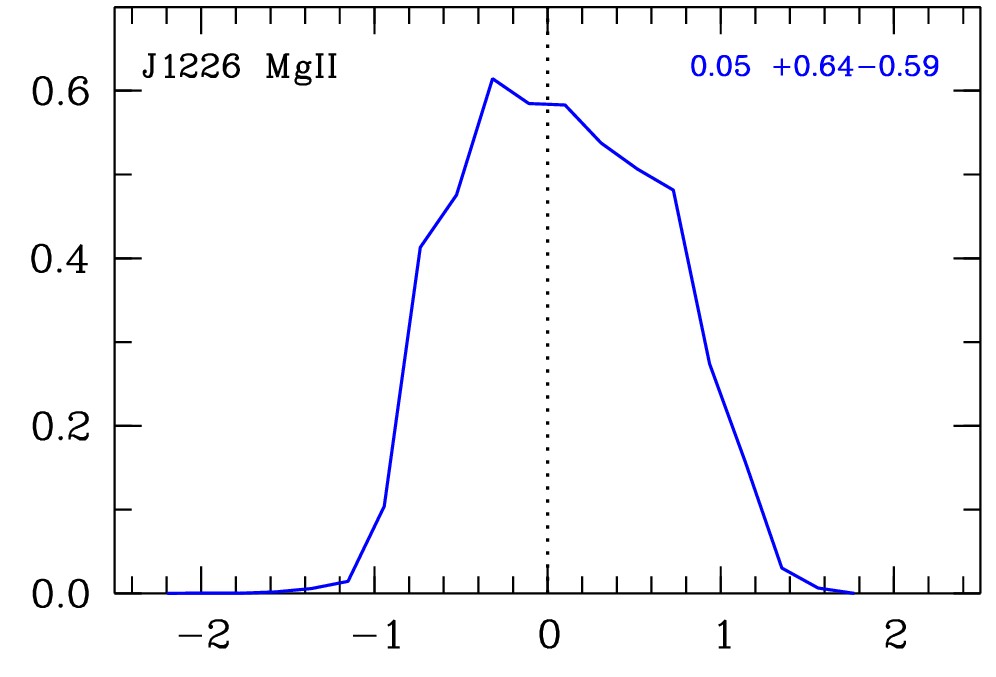}\includegraphics*{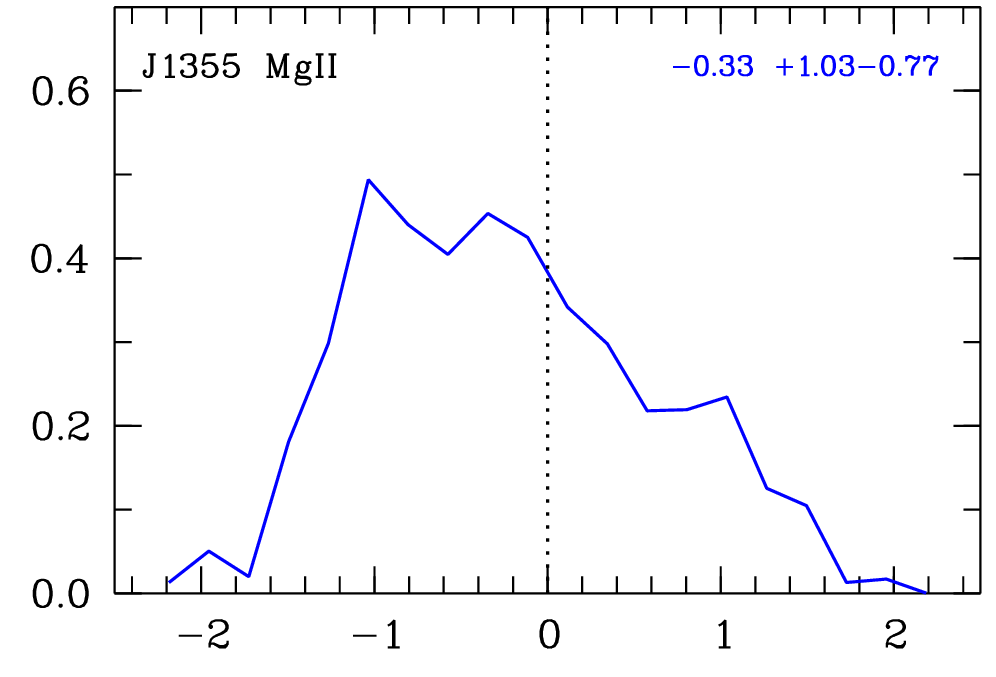}}\\
\resizebox{0.9\hsize}{!}{\includegraphics*{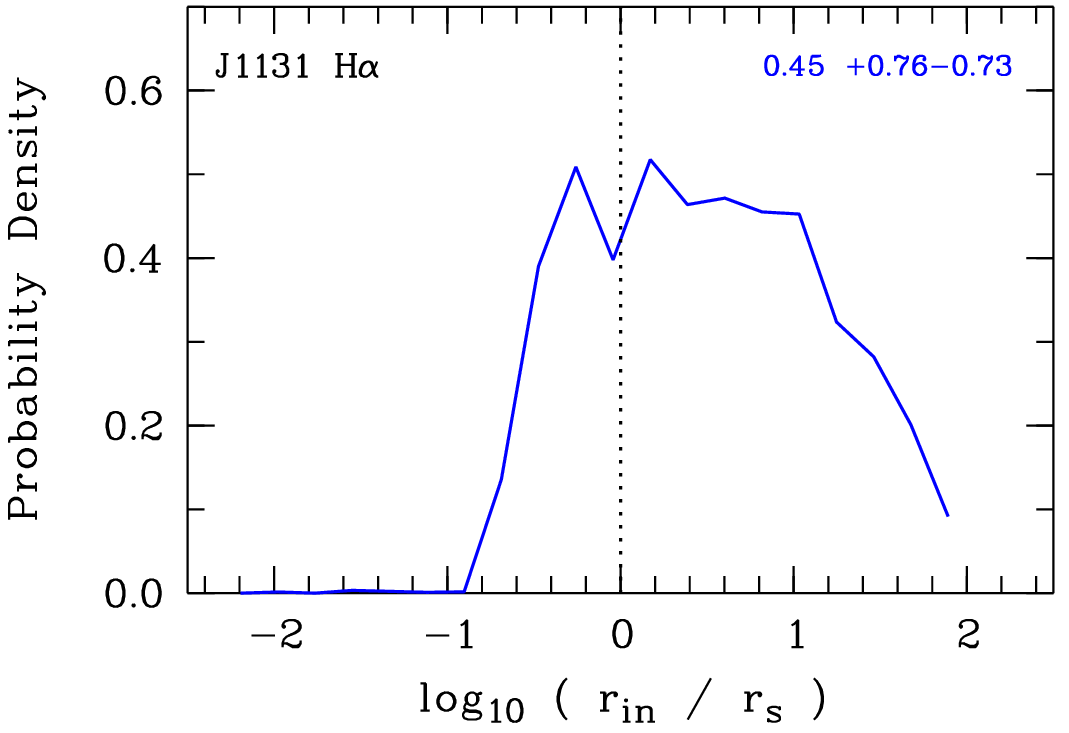}\includegraphics*{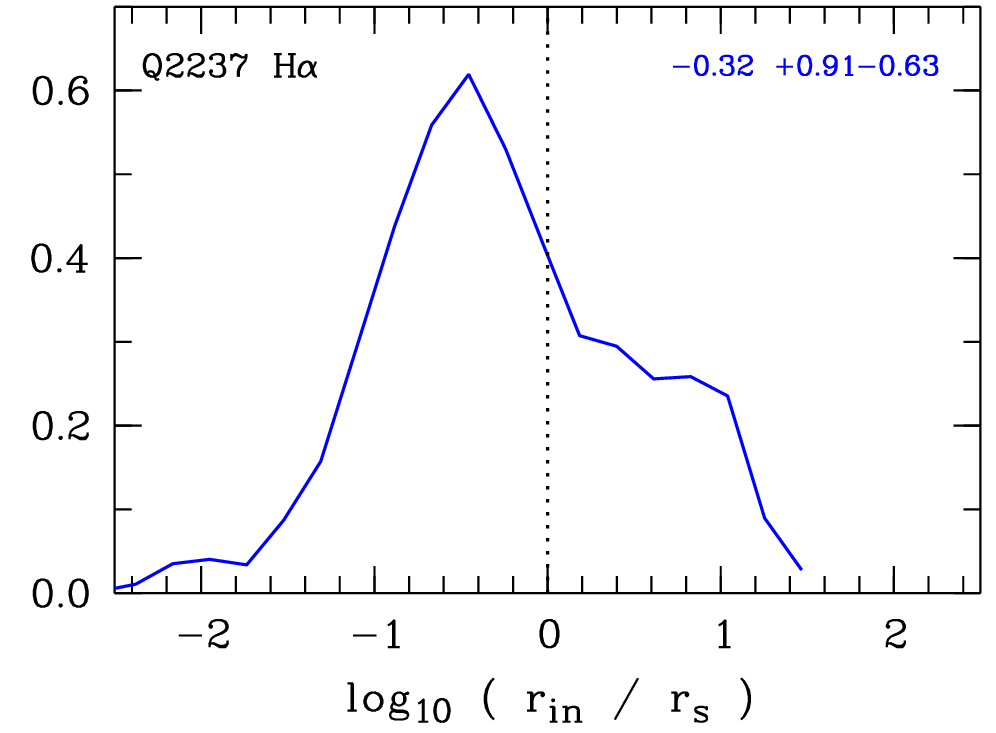}\includegraphics*{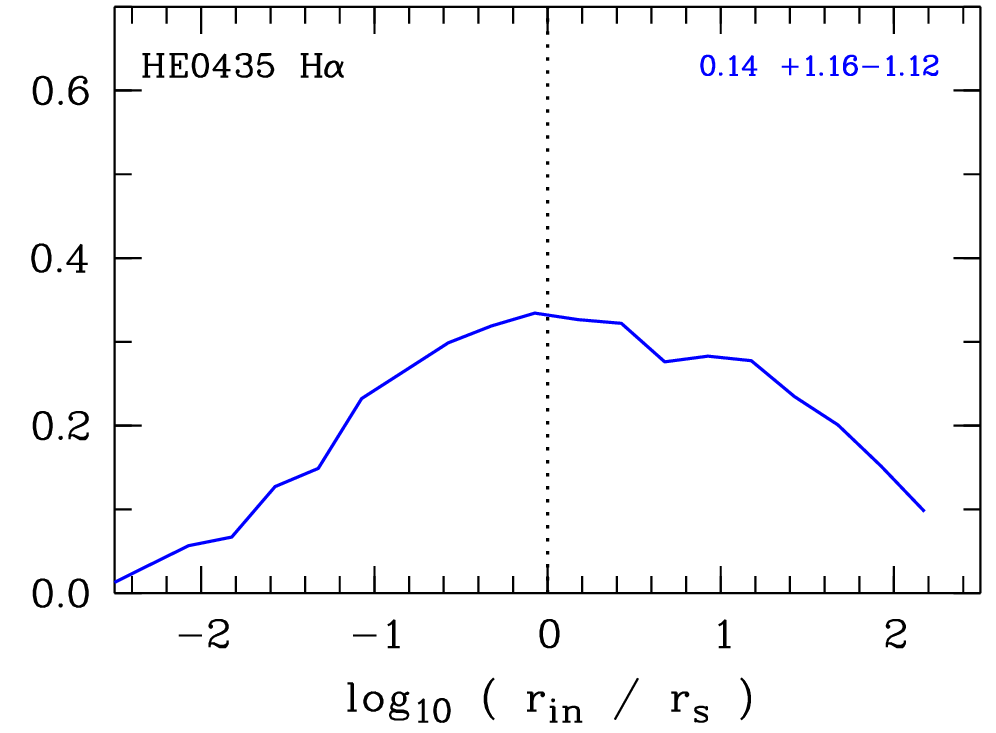}}
\caption{Posterior probability densities of the ratio of the BLR inner radius, $r_{\text{in} \; \text{BLR}}$, over the continuum source outer radius, $r_{\text{s} \; \text{Cont}}$. The object -- line (-- date) combinations are indicated in the left corners. In the right corners, we give the median values of the probability distributions and their uncertainties (see text). The dotted line indicates where  $r_{\text{in} \; \text{BLR}} = r_{\text{s} \; \text{Cont}}$.}
\label{fig:rinrs}
\end{figure*}

\begin{table}[t]
\caption{Targets.}
\label{tab:targets}
\renewcommand{\arraystretch}{1.2}
\centering
\begin{tabular}{lcc}
\hline\hline
 Object & Full quasar name & Redshift $z_s$ \\
\hline
HE0435 & HE~0435$-$1223               & 1.693   \\
J1004  & SDSS J100434.80+411239.2     & 1.734   \\
J1131  & 1RXS~J113155.4$-$123155      & 0.654   \\
J1138  & SDSS~J113803.73$+$031457.7   & 2.438   \\
J1226  & SDSS~J122608.02$-$000602.2   & 1.123   \\ 
J1339  & SDSS~J133907.13$+$131039.6   & 2.231   \\
J1355  & Q1355$-$2257 (CTQ0327)       & 1.370   \\
Q2237  & Q2237$+$0305                 & 1.695   \\
\hline
\end{tabular}
\end{table}

\begin{table*}[t]
\caption{BLR to continuum source radius ratios.}
\label{tab:data}
\renewcommand{\arraystretch}{1.3}
\centering
\begin{tabular}{lccccc}
\hline\hline
Object Line Date & $\log_{10}(L_{\text{bol}})$ & $\log_{10} (r_{\text{in} \; \text{BLR}} / r_{\text{s} \; \text{Cont}})$ & $\log_{10} (r_{\text{1/2} \; \text{BLR}} / r_{\text{1/2} \; \text{Cont}})$ &  $\log_{10} (r_{\text{1/2} \; \text{BLR}})$ & $\log_{10} (r_{\text{1/2} \; \text{Cont}})$\\
\hline
J1004 \ion{C}{iv}       &  45.12$\pm$0.11  &   \hphantom{$-$}$ 0.08^{+0.59}_{-0.78}$ & $0.65^{+0.56}_{-0.61}$  &  $0.45^{+0.23}_{-0.41}$ &                $-0.20^{+0.51}_{-0.46}$ \\  
Q2237 \ion{C}{iv}       &  46.30$\pm$0.30  &                 $-0.11^{+0.87}_{-0.57}$ & $0.36^{+0.77}_{-0.53}$  &  $1.59^{+0.16}_{-0.44}$ &  \hphantom{$-$}$ 1.23^{+0.75}_{-0.29}$ \\  
J1138 \ion{C}{iv}       &  45.40$\pm$0.10  &   \hphantom{$-$}$ 0.13^{+0.95}_{-0.76}$ & $0.59^{+0.98}_{-0.68}$  &  $0.69^{+0.30}_{-0.35}$ &  \hphantom{$-$}$ 0.10^{+0.93}_{-0.58}$ \\  
J1339 \ion{C}{iv} 2014  &  45.88$\pm$0.11  &   \hphantom{$-$}$ 0.24^{+0.73}_{-0.67}$ & $0.99^{+0.70}_{-0.63}$  &  $0.71^{+0.28}_{-0.37}$ &                $-0.28^{+0.64}_{-0.51}$ \\  
J1339 \ion{C}{iv} 2017  &  46.13$\pm$0.11  &   \hphantom{$-$}$ 0.48^{+0.65}_{-0.73}$ & $1.23^{+0.65}_{-0.69}$  &  $0.83^{+0.28}_{-0.26}$ &                $-0.40^{+0.59}_{-0.59}$ \\  
J1339 \ion{Mg}{ii} 2017 &  46.13$\pm$0.11  &   \hphantom{$-$}$ 0.78^{+0.76}_{-0.95}$ & $1.40^{+0.76}_{-0.82}$  &  $1.37^{+0.34}_{-0.45}$ &                $-0.03^{+0.68}_{-0.69}$ \\  
J1131 \ion{Mg}{ii}      &  45.00$\pm$0.10  &   \hphantom{$-$}$ 0.29^{+1.05}_{-0.78}$ & $0.97^{+0.73}_{-0.56}$  &  $0.96^{+0.40}_{-0.32}$ &                $-0.01^{+0.61}_{-0.46}$ \\  
J1226 \ion{Mg}{ii}      &  45.65$\pm$0.15  &   \hphantom{$-$}$ 0.05^{+0.64}_{-0.59}$ & $0.59^{+0.60}_{-0.46}$  &  $0.46^{+0.31}_{-0.26}$ &                $-0.13^{+0.51}_{-0.38}$ \\  
J1355 \ion{Mg}{ii}      &  46.40$\pm$0.10  &                 $-0.33^{+1.03}_{-0.77}$ & $0.50^{+0.85}_{-0.92}$  &  $0.89^{+0.31}_{-0.61}$ &  \hphantom{$-$}$ 0.39^{+0.79}_{-0.69}$ \\  
J1131 H$\alpha$         &  45.00$\pm$0.10  &   \hphantom{$-$}$ 0.45^{+0.76}_{-0.73}$ & $1.06^{+0.70}_{-0.53}$  &  $1.00^{+0.25}_{-0.32}$ &                $-0.06^{+0.65}_{-0.42}$ \\  
Q2237 H$\alpha$         &  46.30$\pm$0.30  &                 $-0.32^{+0.91}_{-0.63}$ & $0.23^{+0.81}_{-0.59}$  &  $1.57^{+0.12}_{-0.42}$ &  \hphantom{$-$}$ 1.34^{+0.80}_{-0.41}$ \\  
HE0435 H$\alpha$        &  45.85$\pm$0.15  &   \hphantom{$-$}$ 0.14^{+1.16}_{-1.12}$ & $0.85^{+1.06}_{-0.96}$  &  $1.18^{+0.50}_{-0.59}$ &  \hphantom{$-$}$ 0.33^{+0.93}_{-0.76}$ \\
\hline
\end{tabular}
\tablefoot{The radius ratios are the median values of the probability densities (see Fig.~\ref{fig:rinrs}). The half-light radii of the BLR are from \citet{2023Hutsemekers,2024Hutsemekers,2024bHutsemekers}. They were measured with an average microlens mass of 0.3 M$_{\odot}$. For Q2237, the half-light radii are taken from Appendix~A of \citet{2024Savic} for consistency with the present work, noting that the time series analysis gives $r_{\text{1/2} \; \text{BLR}} / r_{\text{1/2} \; \text{Cont}}$ a factor of four larger, but still within the uncertainties. The continuum source radii were computed from the BLR to continuum radius ratios. All radii are in light-days. The bolometric luminosities (in erg s$^{-1}$) were computed from the monochromatic luminosities given in \citet{2012Sluse},  \citet{2021Shalyapin}, and  \citet{2023Hutsemekers, 2024Hutsemekers,2024bHutsemekers}, with the bolometric corrections from \citet{2012Sluse}.}
\end{table*}

\begin{figure}[t]
\centering
\resizebox{0.9\hsize}{!}{\includegraphics*{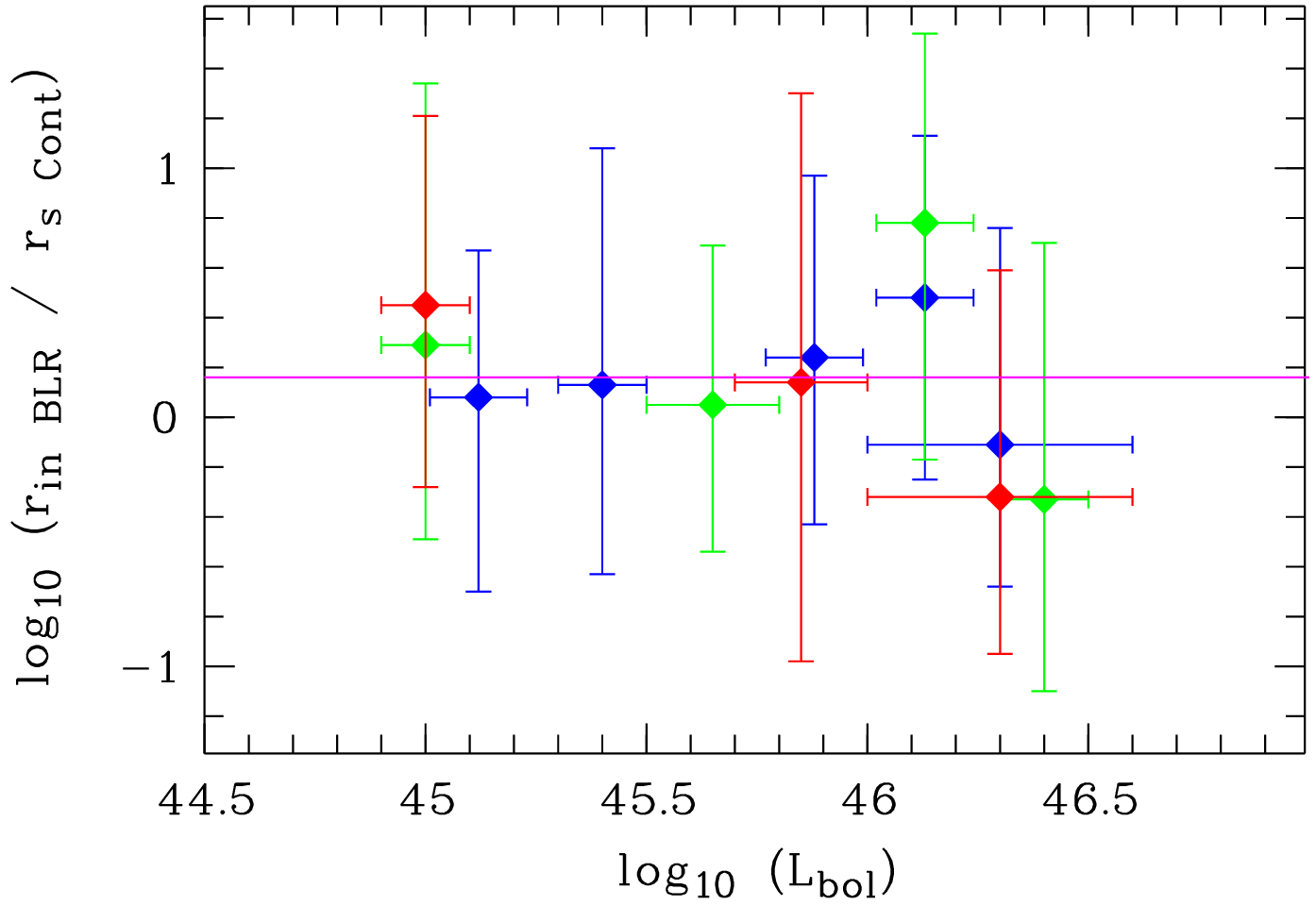}}
\caption{Ratio of the BLR inner to continuum outer radii as a function of the quasar bolometric luminosity. The continuous line shows the sample average. \ion{C}{iv}, \ion{Mg}{ii}, and H$\alpha$ are illustrated in blue, green, and red, respectively.  }
\label{fig:plot1}
\end{figure}

\begin{figure}[t]
\centering
\resizebox{0.9\hsize}{!}{\includegraphics*{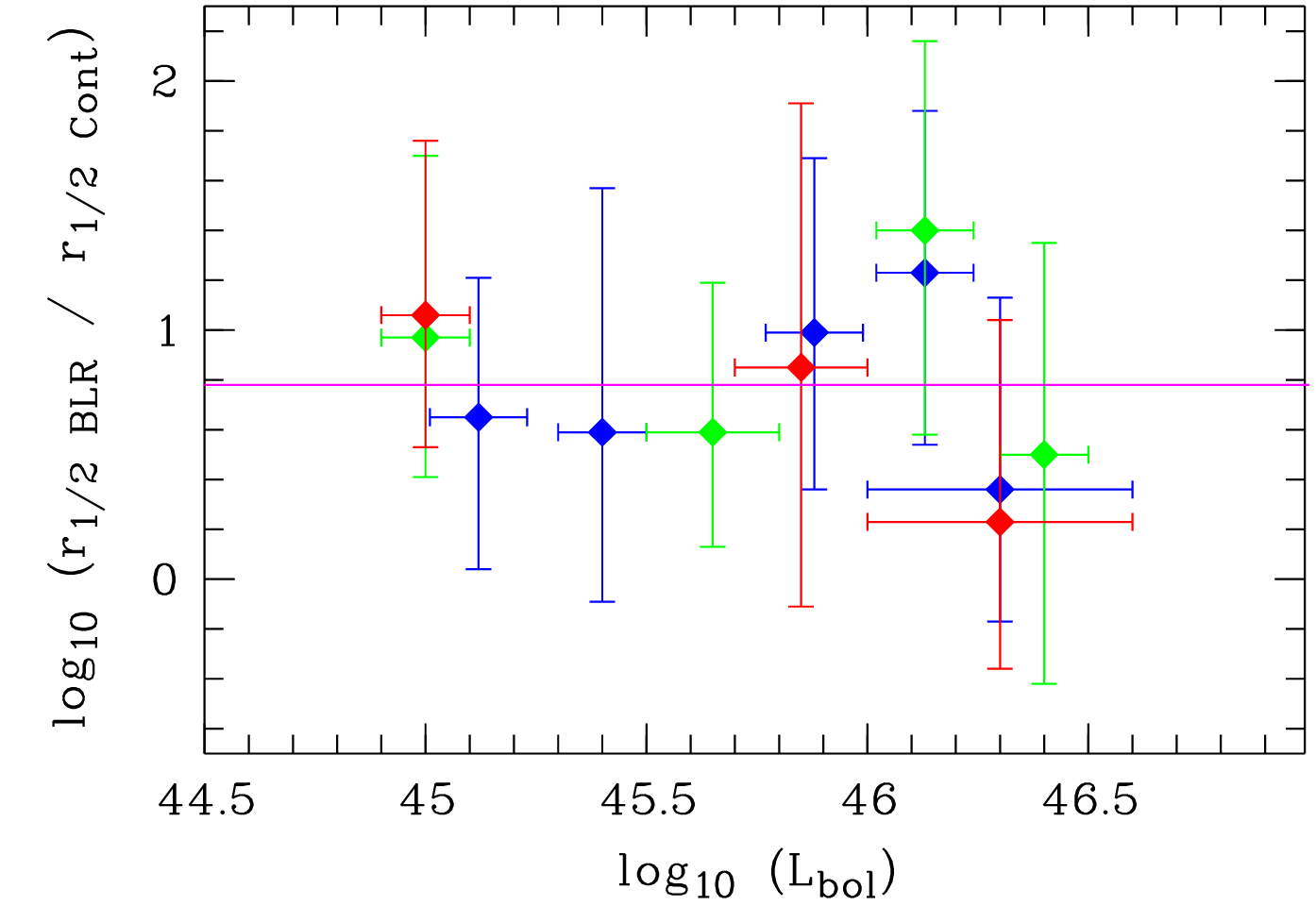}}
\caption{Same as Fig.~\ref{fig:plot1}, but for the ratio of the BLR to continuum half-light radii. }
\label{fig:plot2}
\end{figure}

The relative probabilities of the different combinations of $r_{\text{in}}$ and $r_{\text{s}}$ (hereafter $r_{\text{in} \; \text{BLR}}$ and $r_{\text{s} \; \text{Cont}}$) were obtained by marginalizing the likelihood over all other parameters. Figure~\ref{fig:rinrs} shows the posterior probability densities, uniformly resampled on a logarithmic scale, of the ratio $r_{\text{in} \; \text{BLR}} / r_{\text{s} \; \text{Cont}}$, for the different objects and lines.  The BLR half-light radius, $r_{\text{1/2} \; \text{BLR}}$, which represents the effective size of the BLR probed by microlensing, was computed from $r_{\text{in} \; \text{BLR}}$ for the different BLR models, emissivities, and inclinations \citep[see][]{2021Hutsemekers}, while the half-light radius of the continuum source is simply given by $r_{\text{1/2} \; \text{Cont}} = r_{\text{s} \; \text{Cont}}$ /{\small $\sqrt{2}$} for a uniform disk. It is important to specify that the continuum half-light radius measured with microlensing is essentially independent of the model adopted to describe the source \citep{2005Mortonson}. The relative probabilities of the different combinations of $r_{\text{1/2} \; \text{BLR}}$ and $r_{\text{1/2} \; \text{Cont}}$ were then computed. Since the posterior probability distributions  of $\log_{10} (r_{\text{1/2} \; \text{BLR}} / r_{\text{1/2} \; \text{Cont}})$ are very similar to those obtained for $\log_{10} (r_{\text{in} \; \text{BLR}} / r_{\text{s} \; \text{Cont}})$ (see Fig.~\ref{fig:rinrs}), they are not shown. The median values of the probability distributions are given in Table~\ref{tab:data}. The uncertainties were computed as the equal-tailed credible intervals that enclose a posterior probability of 68\%. The median values are plotted in Figs.~\ref{fig:plot1} and~\ref{fig:plot2} as a function of the quasar bolometric luminosity, $L_{\text{bol}}$,  for the sample of 12 [object,line,date] combinations. We emphasize that the continuum sizes were measured at the wavelengths of the emission lines; that is, 1550~\AA, 2800~\AA, and 6563~\AA\ in the quasar rest frame. It is also important to note that while the BLR and continuum source radii depend on the adopted average microlens mass (usually 0.3 M$_{\odot}$), their ratio is completely independent of it.

It is immediately clear from Fig.~\ref{fig:rinrs}  that the majority of the $\log_{10} (r_{\text{in} \; \text{BLR}} / r_{\text{s} \; \text{Cont}})$ probability distributions are centered on $r_{\text{in} \; \text{BLR}} \simeq r_{\text{s} \; \text{Cont}}$. Figure~\ref{fig:plot1} shows that the median values are independent of $L_{\text{bol}}$.  The average of the full sample (12 object+line+date combinations) is $\log_{10} (r_{\text{in} \; \text{BLR}} / r_{\text{s} \; \text{Cont}}) = 0.16 \pm 0.32$, where the quoted uncertainty is the standard deviation. The inner radius of the BLR is thus equal to the outer radius of the continuum source within a factor of two, with possible overlaps. The distributions and median values of $\log_{10} (r_{\text{1/2} \; \text{BLR}} / r_{\text{1/2} \; \text{Cont}})$ show exactly the same behavior, with an offset (Fig.~\ref{fig:plot2}). The average of the full sample is $\log_{10} (r_{\text{1/2} \; \text{BLR}} / r_{\text{1/2} \; \text{Cont}}) = 0.78 \pm 0.36$, which means that the half-light radius of the BLR is on average a factor of six larger than the half-light radius of the continuum source. It is noteworthy that there is no significant difference between the high- and low-ionization lines, which also means between the continua measured at different wavelengths.\footnote{Only \ion{Mg}{ii} in J1339 seems to have a slightly higher $r_{\text{in} \; \text{BLR}} / r_{\text{s} \; \text{Cont}}$ ratio, although the difference is not significant given the large error bars.} Systematic errors affecting the different lines, if any, would not produce a constant $r_{\text{in} \; \text{BLR}} / r_{\text{s} \; \text{Cont}}$ ratio, but rather scramble it. 

The fact that the ratio of the BLR to continuum size is constant, within the uncertainties, for different quasar luminosities suggests that the BLR and continuum source sizes should change in parallel. We then computed the continuum source radii from the radius ratios we have measured and the BLR radii reported in our previous work (Table~\ref{tab:data}). The BLR radii are plotted against the continuum source radii in Fig.~\ref{fig:plot4}, where a clear correlation can be observed.\footnote{\citet{2021Fian,2023Fian} estimated the BLR and the continuum source radii using microlensing in the lensed quasar Q0957$+$561, which is not in our sample. They found $r_{\text{1/2} \; \text{BLR}} \simeq$ 50 light-days for the \ion{Mg}{ii} BLR and  $r_{\text{1/2} \; \text{Cont}} \simeq$ 18 light-days for the continuum at 2558~\AA. These values follow and support our results (assuming the size of the continuum source at 2800~\AA\ similar to its size at 2558~\AA).}

\section{Discussion}
\label{sec:discussion}

\begin{figure}[t]
\centering
\resizebox{0.9\hsize}{!}{\includegraphics*{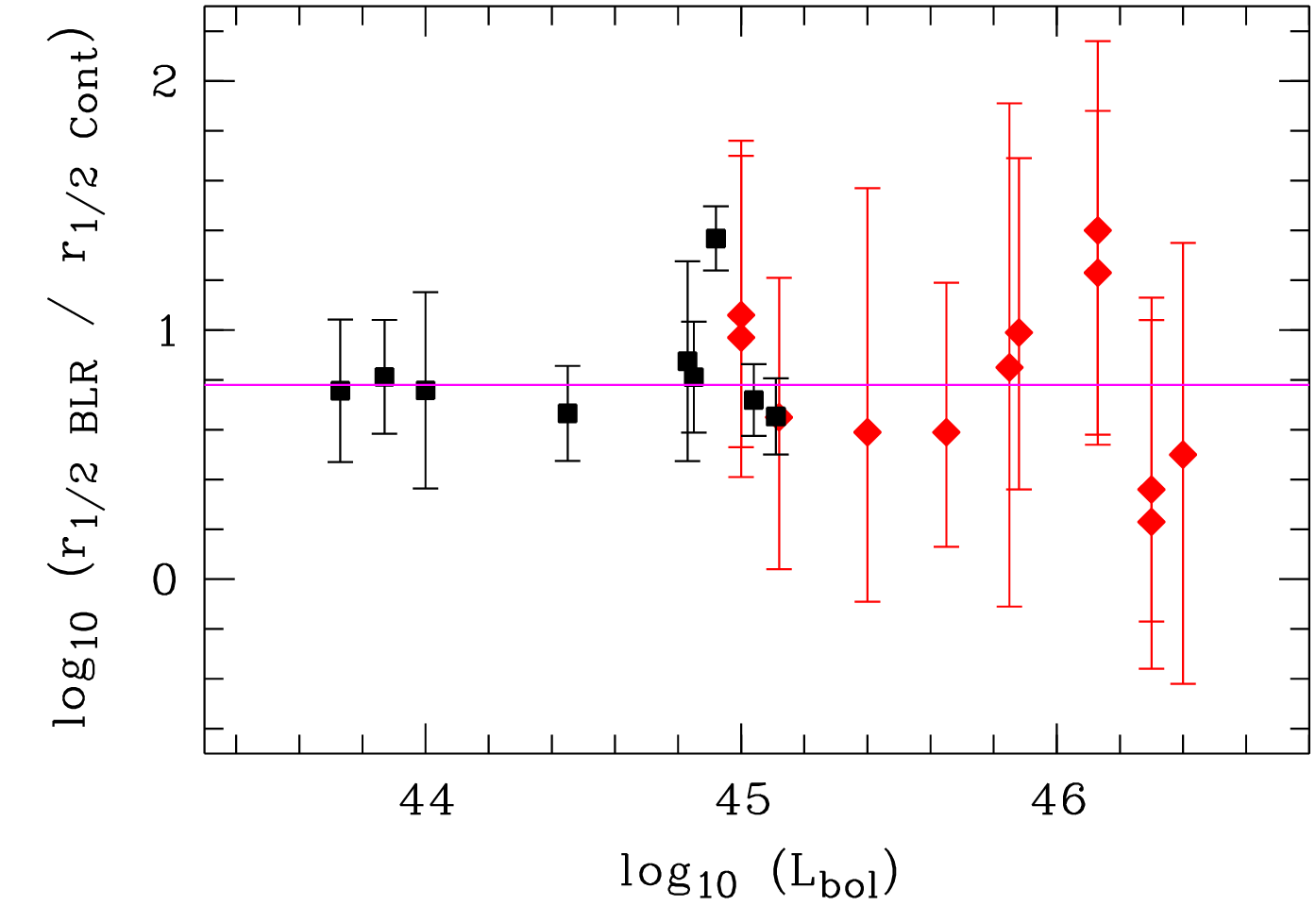}}
\caption{Ratio of the BLR to continuum sizes as a function of the quasar bolometric luminosity.  The RM H$\beta$ to 5100~\AA\ time lag ratios from \citet{2022Netzer} are in black (squares). Our microlensing half-light radius ratios are shown in red (diamonds). The continuous line represents the average of our sample.}
\label{fig:plot3}
\end{figure}

\begin{figure}[t]
\centering
\resizebox{0.9\hsize}{!}{\includegraphics*{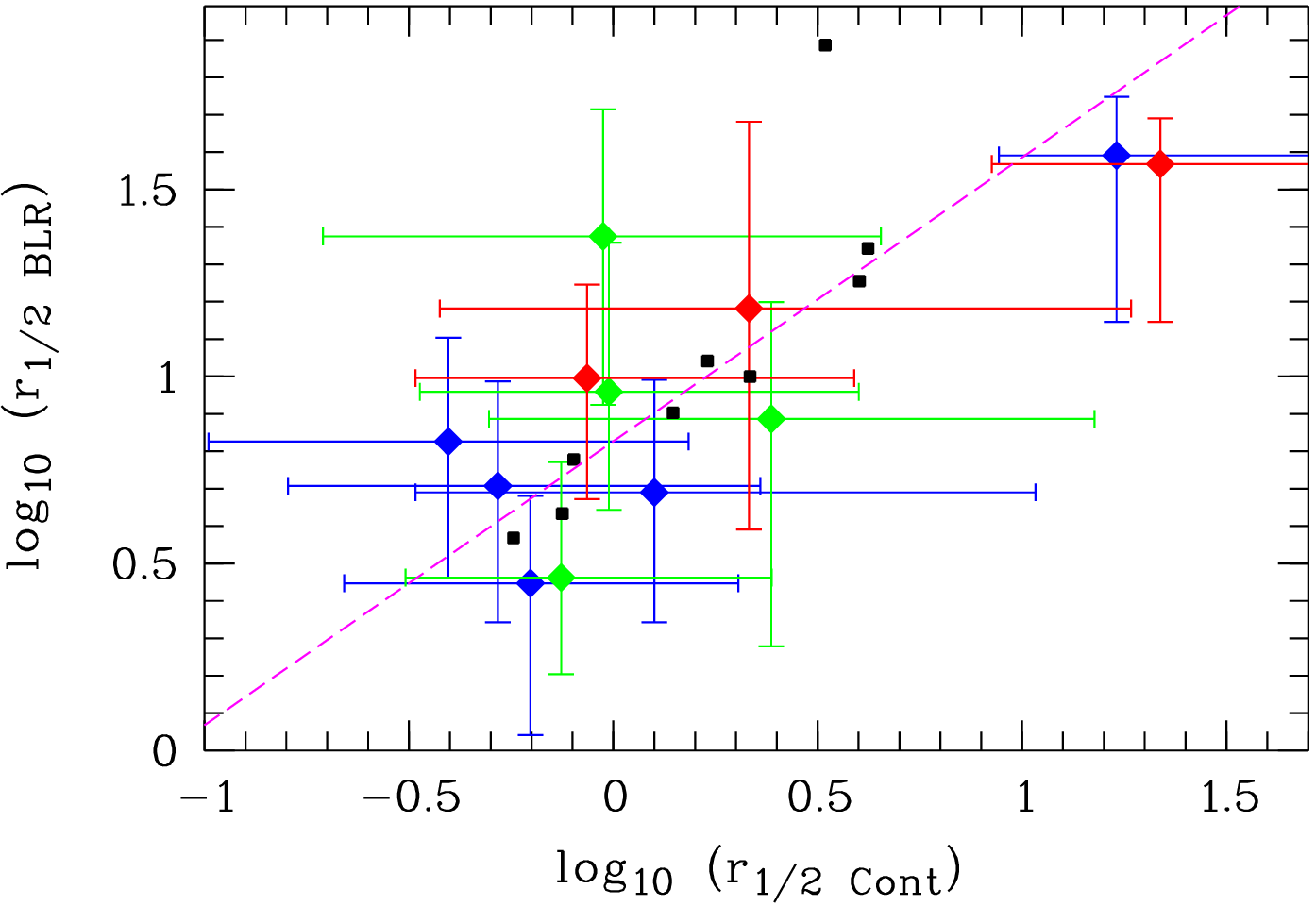}}
\caption{BLR half-light radius as a function of the continuum source half-light radius. \ion{C}{iv}, \ion{Mg}{ii}, and H$\alpha$ are illustrated in blue, green, and red, respectively. The additional small black squares show the RM measurements considered by \citet{2022Netzer}. The dashed line represents the \citet{2023Wang} relation, $\log_{10} (r_{\text{1/2} \; \text{BLR}}) = 0.759 \,  (\log_{10} (r_{\text{1/2} \; \text{Cont}}) - \rho) + 1.434$, with $\rho$ = 0.8 instead of 0.5.}
\label{fig:plot4}
\end{figure}

\begin{figure}[t]
\centering
\resizebox{0.9\hsize}{!}{\includegraphics*{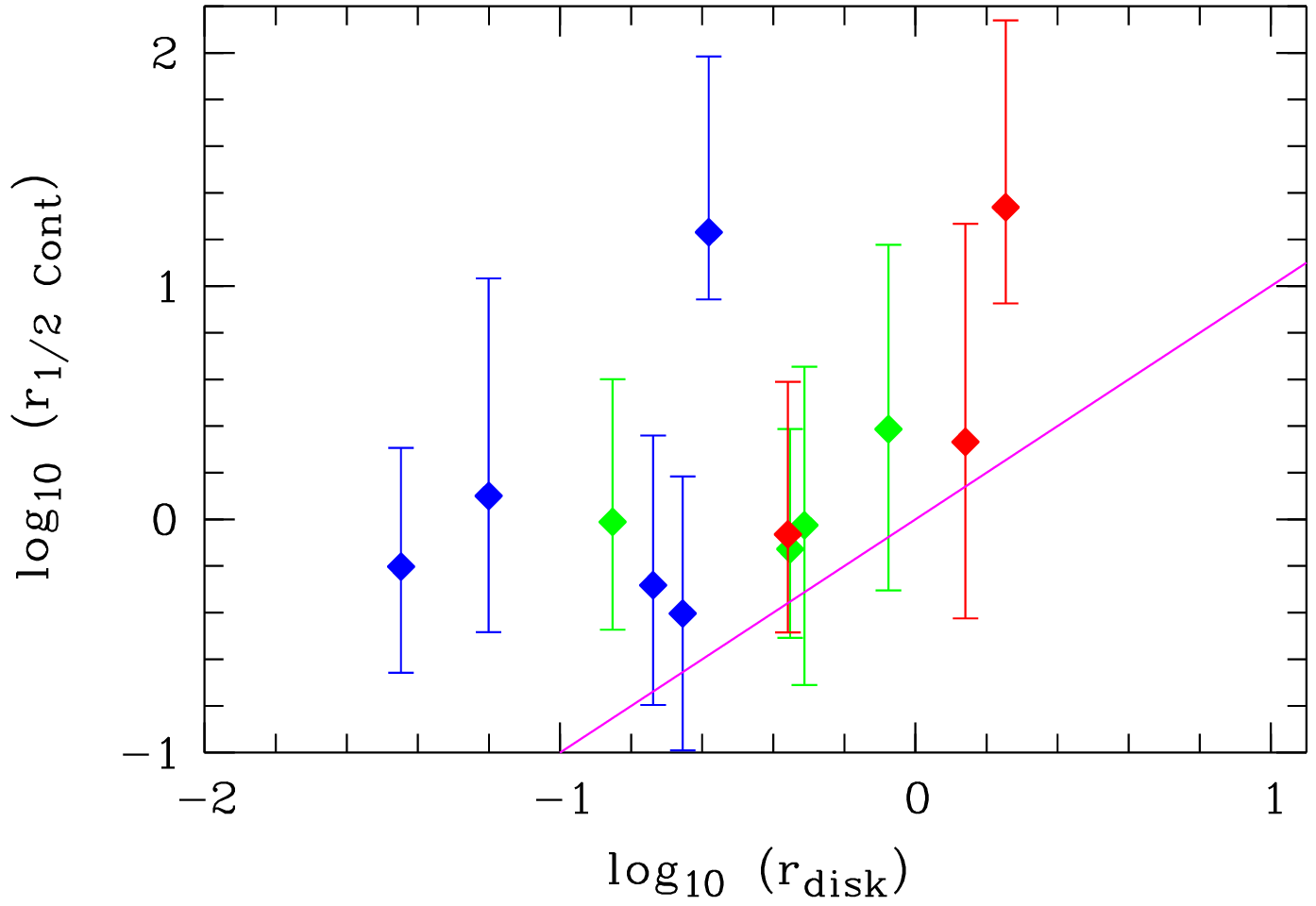}}
\caption{Continuum source half-light radii from this work versus the Shakura-Sunyaev disk radii expected at the observed continuum wavelengths. The latter radii were computed using Eq.~3 of \citet{2011Mosquerab} with an accretion efficiency of $\eta = 0.1$, and black-hole masses from \citet{2012Sluse}, \citet{2021Shalyapin}, and \citet{2023Hutsemekers}. \ion{C}{iv}, \ion{Mg}{ii}, and H$\alpha$ are illustrated in blue, green, and red, respectively. The continuous line indicates where $\log_{10} (r_{\text{1/2} \; \text{Cont}})$ = $\log_{10} (r_{\text{disk}})$.  }
\label{fig:plot5}
\end{figure}

These results independently confirm the RM findings that the optical continuum source is generally comparable in size to the inner BLR \citep{2016Fausnaugh,2021Kara}. Moreover, the factor of six between the BLR and the continuum source half-light radii is in excellent agreement with the RM H$\beta$ to 5100~\AA\ time lag ratios reported by \citet{2022Netzer}.  In this sample, the continuum time lags were measured between the ionizing far-UV and the 5100~\AA\ continua, providing a good estimate of the size of the continuum source at 5100~\AA. Our measurements thus extend the RM results, in particular the luminosity independence of the BLR to continuum size ratio, to higher-redshift, higher-luminosity quasars (Fig.~\ref{fig:plot3}). The correlation between the BLR and the UV-optical continuum source radii, which is also in excellent agreement with the RM time lags of \citet{2022Netzer} (Fig.~\ref{fig:plot4}), supports the idea that the dominant contribution to the UV-optical continuum may originate from the BLR itself, with both the UV-optical continuum and the BELs responding similarly to the ionizing continuum coming from a more compact source.\footnote{\citet{2023Wang} report a relation between the H$\beta$ BLR and the 5100~\AA\ continuum source radii, based on RM measurements that include those of \citet{2022Netzer}. With an offset of 0.3~dex toward larger continuum sizes, this relation is in good agreement with both our measurements and those of \citet{2022Netzer} (Fig.~\ref{fig:plot4}). The offset could be due to the fact that the \citet{2023Wang} sample also includes objects for which only optical time lags were available, which could bias the continuum source radii toward smaller values. We emphasize that microlensing directly measures sizes, and thus cannot be affected by such a bias.} Remarkably, the correlation holds for lines of different ionization and their underlying continua at different wavelengths, at least for the objects of our sample. In agreement with previous work \citep{2010Morgan,2011Blackburne,2020Cornachione}, we also found that the half-light radius of the continuum source is, on average, a factor of four larger than the size expected from the standard Shakura-Sunyaev accretion disk (Fig.~\ref{fig:plot5}). 

Various photoionization models \citep{2001Korista,2018Lawther,2019Chelouche,2019Korista,2022Netzer} have shown that a diffuse continuum emitted in the BLR over the full UV-optical wavelength range can quantitatively reproduce the observed RM continuum time lags, eclipsing the continuum previously thought to be emitted by a standard accretion disk. A similar conclusion was reached by simulating microlensing-based continuum size measurements with a diffuse, extended continuum source \citep{2023Fian}. While RM probes the time lags between the quasar variable components, microlensing, on the other hand, is sensitive to the size of the light-emitting regions.

Although they all agree on the importance of the diffuse continuum coming from the BLR, the models differ in their assumptions about the BLR physics and photoionization. In particular, they can predict different contributions of the diffuse continuum to the total UV-optical continuum, and different wavelength dependences of this contribution. The fact that the correlation between the BLR and the continuum source radii is observed at both 1550~\AA\ and 6563~\AA\ (Fig.~\ref{fig:plot4}) suggests that the diffuse continuum contribution could be significant over at least the wavelength range 1550 -- 6563~\AA. If confirmed with more data and/or smaller uncertainties, such a result could put constraints on the BLR physics. 

An extended, non-microlensed continuum source was also found in two lensed broad absorption line quasars, thanks to the differential absorption of the microlensed and non-microlensed continua \citep{2015Sluse,2020Hutsemekers}. For one quasar, spectropolarimetric observations suggested that the diffuse continuum can be attributed to scattering from an extended region located along the polar axis \citep{2015Hutsemekers}. If the extended continuum source is due to scattering, a correlation between the BLR and the continuum source radii would not be expected. However, the scattering interpretation could only apply to broad absorption line quasars that show prominent winds, and are, on average, more polarized than normal quasars \citep{1984Stockman,1998Hutsemekers}.

\section{Summary and conclusions}
\label{sec:conclu}

We have reported measurements of the ratio of the BLR radius to the UV-optical continuum source radius, for eight lensed quasars and different emission lines (\ion{C}{iv}, \ion{Mg}{ii}, and H$\alpha$). These measurements are based on the observed microlensing-induced BEL profile distortions and underlying continuum magnifications, which were simultaneously compared with microlensing simulations. We found that, on average, and independently of the line ionization level and the wavelength of the continuum, the inner radius of the BLR starts at the end of the UV-optical continuum source. We found that the half-light radius of the BLR is, on average, a factor of six larger than the half-light radius of the continuum source, independently of the quasar bolometric luminosity. We also found a correlation between the BLR radius and the continuum source radius, supporting the idea that the dominant contribution to the UV-optical continuum comes from the BLR itself. This correlation involves lines of different ionization with their underlying continua at different wavelengths, suggesting that the contribution of the diffuse UV-optical continuum may be significant over a wide range of wavelengths, although more data with smaller uncertainties would be needed to confirm this. We emphasize that these results are independent of the mass of the microlens. By directly probing the size of light-emitting regions rather than time lags of variable components, our microlensing analysis independently confirms previous RM results, and extends them to higher-redshift, higher-luminosity quasars.

Although currently based on a small sample with rather large uncertainties, our results open interesting perspectives on understanding the nature of the diffuse, extended continuum source in quasars, in relation to the BLR. They complement the results from RM by probing different aspects of the quasar inner regions, the size of the light emitters versus time lags, and for quasars at higher redshifts.

\begin{acknowledgements}
D.H. thanks Mohamad Naddaf for useful discussions on BLR models, and acknowledges support from the Fonds de la Recherche Scientifique - FNRS (Belgium) under grants PDR~T.0116.21 and No 4.4503.19.
\end{acknowledgements}

\bibliographystyle{aa}
\bibliography{references}

\end{document}